\documentclass[12pt,preprint]{aastex}

\def\kms{\relax \ifmmode {\,\rm km~s}^{-1}\else \,km~s$^{-1}$\fi}
\def\cm-3{\relax \ifmmode {\,\rm cm}^{-3}\else \,cm$^{-3}$\fi}
\def\deg{\hbox{$^\circ$}}
\def\arcsec{\hbox{$^{\prime\prime}$}}
\def\h13cn{H$^{13}$CN}
\def\h13co+{H$^{13}$CO$^+$}

\begin{document}

\title{
High Angular Resolution Measurement  of   Ion and Neutral Spectra as a Probe
of the Magnetic Field Structure in DR21(OH)}
       
\author{Shih-Ping Lai\altaffilmark{1}, T. Velusamy, and W. D. Langer\\
{\it slai@astro.umd.edu, velu@jpl.nasa.gov, william.langer@jpl.nasa.gov}}
\affil{Jet Propulsion Laboratory, California Institute of Technology, MS 
169-506, 4800 Oak Grove Dr., Pasadena, CA 91109}
\altaffiltext{1}{Current address: Department of Astronomy, University of 
Maryland, College Park, MD 20742-2421}

\begin{abstract} 
It has been suggested that under average interstellar field
strengths the cyclotron interaction between ions and magnetic fields
is strong enough to narrow the linewidth and suppress the line wings
in the ion spectra.    We present evidence for the 
cyclotron interaction effect at arcsec scale on the 
velocity dispersions in the spectra of ion/neutral molecular species     
in DR21(OH) observed with the OVRO-MMA.  
Using a spatial resolution $\sim$ 3 times higher than  previous 
CSO observations  by Houde et al.\ (2002), 
we show that H$^{13}$CO$^+$ and H$^{13}$CN are coexistent at the scale
of our observations (6\arcsec).
In the eastern parts of the DR21(OH) core where the dynamics is simple,
the ion linewidths are indeed narrower than the neutral linewidths with an
average ion-to-neutral linewidth ratio of 0.82$\pm$0.04.
We use our results, along with the existing Zeeman and dust/CO polarization
data on small scales, to derive the 3-D magnetic field structure. 
We obtain a field strength of $0.44\pm$0.12 mG with inclination of 36\deg\ 
to the line of sight, directed toward the observer, and a position angle of 
$-$75\deg\ in the plane of the sky.  
With the full magnetic field strength derived here,
we are able to conclude that the MM1 core of DR21(OH) is magnetically 
supercritical; although turbulence provides the dominant support.

\end{abstract}

\section{Introduction}

Measurements of magnetic fields in the environment of star forming regions
are mostly carried out with observations 
of Zeeman splitting (Crutcher 1999),
the polarimetry of dust continuum (Dotson et al.\ 2000;
Lai et al.\ 2001, 2002), and the polarimetry of CO emission
(Girart et al.\ 1999; Lai et al.\ 2003).
The Zeeman observations only give the field {\it strength} along the line of
sight, and the polarimetry of dust continuum and CO emission
yield the field {\it direction} in the plane of sky.
As a result, these observations do not determine
the full strength and orientation of the magnetic fields in 3-D.  
To characterize the magnetic field environment,
we need to measure directly the magnetic field strength in the plane of sky 
$B_p$.   
Without direct measurement of the full field strengths,
the role magnetic fields play in star formation can
only be assessed using indirect estimates of $B_p$ 
(the Chandrasekhar-Fermi method as used by Lai et al.\ 2001)
and some statistical arguments (Crutcher 1999).

Houde et al.\ (2002) proposed an interesting
new approach to probe the orientation of magnetic fields. 
Houde et al.\ (2000a,b) show that in weakly ionized plasma,
even under weak fields ($\sim 10~\mu$G),
the cyclotron interaction between the ions and the field can
reduce the linewidth and suppress the high velocity wings in the ion
spectra when compared to coexistent neutral spectra.
Their HCN and HCO$^+$ observations with the Caltech Submillimeter 
Observatory (CSO) show evidence of these predicted features.
Because the ion-to-neutral linewidth ratio depends largely on the
inclination angle of the magnetic fields (no cyclotron interaction
along the field lines),
the full strength and orientation of magnetic fields can
be obtained by combining the observations of Zeeman splitting,
polarimetry of dust continuum and/or CO emission,
and the ion-to-neutral linewidth ratio.
Houde et al.\ (2002) have constructed the 3-D field structure of M17
using this method.

However, it is difficult to interpret 
the differences in the velocity dispersions of the ion/neutral 
spectra unambiguously as being caused by the cyclotron interaction,
if the beam is large enough to include the spectral line 
emissions from physically and chemically different components.
The intrinsic difference between the velocity structure of
the ions and neutrals will also contribute to 
the ion-to-neutral linewidth ratio.
For example, if a velocity gradient exists in a large region with HCN,
but HCO$^+$ only coexists with HCN in a small part of this region,
the HCN linewidth is naturally larger than the HCO$^+$ linewidth.
Therefore, it is important to obtain high spatial resolution
spectral line maps to resolve the regions with widely 
different physical and chemical characteristics.

In this paper, we present the comparison between
H$^{13}$CN and H$^{13}$CO$^+$ spectra in DR21(OH) observed with 
the millimeter array of the Owens Valley Radio Observatory (OVRO).
DR21(OH) is an excellent object to study   the magnetic
field structure, because it is in the early stage of massive star formation.
DR21(OH) is one of the few sources with measured average line-of-sight 
magnetic field strength from CN Zeeman detections (Crutcher et al.\ 1999)  
and field directions in the plane of sky   mapped with BIMA
using both dust and CO polarimetry (Lai et al.\ 2003).
Deriving the full strength and 3-D  structure requires knowledge of   the 
inclination angle of the magnetic field to 
the line of sight, which can be obtained by measuring the  
ion-to-neutral linewidth ratio.

\section{Observations and Results}

The observations were made with the OVRO Millimeter Array
at 3mm in the Compact, Low, and High resolution array configurations 
between  October 2002 and May 2003.  We obtained simultaneously spectral line 
images  of HCN \& HCO$^+$ and  H$^{13}$CN \& H$^{13}$CO$^+$;   
$^{13}$CS and N$_2$H$^+$ were observed in the opposite sidebands.
The synthesized beam is 6.4\arcsec$\times$6.9\arcsec\ for natural weighting
and spectral resolutions are 0.434 \kms.
Among these images, the HCN and HCO$^+$ spectra show features of   
optically thick emission with strong self-absorption;
N$_2$H$^+$ has a very different spatial distribution
compared to other species;  H$^{13}$CN, H$^{13}$CO$^+$, and $^{13}$CS
have a similar spatial distribution, but $^{13}$CS has a different
velocity structure from the other two.  Therefore, 
H$^{13}$CN and H$^{13}$CO$^+$ are the only neutral/ion pair that
can be used to examine the cyclotron effect in our data.

Figure 1 shows the 3mm continuum and the integrated flux of
H$^{13}$CN and H$^{13}$CO$^+$.  The two compact cores in DR21(OH),
MM1 and MM2, are resolved.  H$^{13}$CN and H$^{13}$CO$^+$ both
trace the dense cores very well.  H$^{13}$CO$^+$ is slightly more
extended than H$^{13}$CN, which is probably due to the lower
critical density of H$^{13}$CO$^+$.  Therefore, 
H$^{13}$CN and H$^{13}$CO$^+$ seem to occupy the same volume.  
Figure 2 shows the normalized H$^{13}$CN and H$^{13}$CO$^+$ 
spectra overlaid on their flux ratio map.  
Note that H$^{13}$CN has three hyperfine lines with intensity ratios 
of 1:5:3 for the optically thin case.  The spectra show that DR21(OH) 
contains two dominant velocity components -- one at $\sim -5$\kms\ and
the other at $\sim -1$\kms, which are likely to be
associated with MM1 and MM2, respectively.
At the positions between MM1 and MM2, the spectra are more complex
and may contain  
an absorption feature at $\sim -$2.5\kms.  Overall, the spectra
of H$^{13}$CN and H$^{13}$CO$^+$ trace each other well, which is strong 
evidence for the coexistence of H$^{13}$CN and H$^{13}$CO$^+$, 
as also suggested by their spatial intensity correlation.  

A difference in the  ion/neutral linewidth   can arise in some cases from 
chemical 
inhomogeniety, especially, if multiple velocity features are present. In Figure 
2 we show the distribution of the H$^{13}$CN-to-H$^{13}$CO$^+$ 
flux ratio (in greyscale) which is a measure of the chemical 
differentiation in the core. We can  safely 
discard the effects of chemical differentiation by excluding regions where the 
flux ratio changes 
appreciably over the small spatial scale  sizes of the OVRO beam. However, 
chemical 
differentiation alone will not always cause linewidth differences. The velocity 
structure of the region also needs to be considered.  
For example, if the velocity structure is uniform across the cores 
such that the shape and the central velocity of the spectra stay the same, 
any local variation in the ion-to-neutral flux ratio will result in a weighting 
factor that  affects only the line intensities, not the   
 ion-to-neutral linewidth ratios
for the whole core. 
However, in DR21(OH)   at least two velocity components are detected  and 
therefore,
it is critical to isolate the regions with different dynamics when
interpreting the ion-to-neutral linewidth ratio as the cyclotron effect.
We see a large variation in their flux ratio (by factors of 0.5 to 4) over the 
entire region, 
but it appears to be smooth except near the southwest parts of the core.
  In the eastern parts of the 
core the variation in the flux ratio is  small, 
and the fully resolved velocity structure is dominated by a single feature. 
Thus, in this region of DR21(OH), we are able to  
minimize  any contribution to the ion/neutral linewidth differences arising 
from   chemical inhomogeneity. In our analysis we do not include the southwest 
region where rapid and significant variation of ion-to-neutral flux ratio is 
observed.
Furthermore, the  ion/neutral spectra in the southwest region have low S/N.

In the eastern part of MM1 where the $-$5\kms\ component dominates and
the spectral line shapes are better defined, 
we can see that H$^{13}$CO$^+$ is slightly narrower 
than H$^{13}$CN.  To eliminate the complication of the H$^{13}$CN
hyperfine lines, we simulate H$^{13}$CN spectra with all three hyperfine 
lines using the H$^{13}$CO$^+$ as a `template'. 
For the case of a single velocity component,
if the linewidth of H$^{13}$CN is smaller than 
half of the velocity separations between hyperfine lines,
the difference spectrum between the observed H$^{13}$CN and 
that simulated from H$^{13}$CO$^+$ is 
largely free from the confusion caused by the hyperfine components
and thus is a useful measure of the cyclotron interaction.
If indeed H$^{13}$CN has larger linewidths than H$^{13}$CO$^+$,
the difference spectra will have two positive peaks near 
the wings of the line profile as demonstrated in 
Figure 2d. As the line shapes of H$^{13}$CN and H$^{13}$CO$^+$
are approximately Gaussians, the integration of the difference spectra over 
the extent of the main hyperfine component gives a measure of  
the difference of the dispersions in the spectra.
\begin{equation}
\int{(I_n - I_i)}{dv} = \sqrt{2\pi} (\sigma_n - \sigma_i),
\end{equation}
\noindent where $I_n$ and $I_i$ are the normalized H$^{13}$CN and 
H$^{13}$CO$^+$ spectra and $\sigma_n$ and $\sigma_i$
are the dispersions of H$^{13}$CN and H$^{13}$CO$^+$, respectively.
Figure 3a shows the map of $(\sigma_n - \sigma_i)$ which is calculated by 
integrating the difference spectra 
over the $-5$\kms\ component ($v_{LSR}=-9$ to $-2.5$\kms). 
These $(\sigma_n - \sigma_i)$ estimated using Eq(1) are in good
agreement (within 10\%) with those obtained by fitting Gaussians to
the observed H$^{13}$CN and H$^{13}$CO$^+$ line profiles at (+10\arcsec, 
+5\arcsec).
The signal-to-noise ratio (S/N) of $(\sigma_n - \sigma_i)$
is presented  in Figure 3b. 
We use $(\sigma_n - \sigma_i)$ and $\sigma_i$ 
measured from H$^{13}$CO$^+$ 
spectra to derive $(\sigma_i/\sigma_n)$ in Figure 3c.
It is evident from the results shown in Figure 3 that in the eastern 
side of MM1 the ion linewidth is smaller than the neutral linewidth.
We do not present similar analysis for the $-1$\kms\ component,
because of the low S/N of $(\sigma_n - \sigma_i)$.

\section{Discussion}

The interferometer observations with OVRO have the advantage to resolve 
the velocity and chemical structures in the  DR21(OH) core with high spatial 
resolution.  In the eastern parts of DR21(OH)
our observations show that H$^{13}$CN and H$^{13}$CO$^+$ coexist over ~15\arcsec 
$\times$10\arcsec\ in a single velocity feature
  and confirm  that
the ion line is indeed narrower than the neutral line.  
We use the ion-to-neutral linewidth ratio $(\sigma_i/\sigma_n)$ shown in Figure 
3c
to infer the inclination angle of the magnetic field to 
the line of sight ($\alpha$) according to Equation (11)
of Houde et al.\ (2002).  To derive the value of ($\alpha$) using the 
ion/neutral linwidth ratio we require   knowledge of the gas motions within 
the core.
Houde et al.\ consider the cases with neutral flow 
collimation widths, $\Delta\theta$, ($\Delta\theta=90$\deg\ means no 
collimation as in a  purely turbulent case).  For a given value of 
$(\sigma_i/\sigma_n)$  a range of values for $\alpha$ are possible depending 
on the values for flow collimation widths (see Figure 4 in Houde et al.). For 
example the
average ion-to-neutral linewidth ratio, $(\sigma_i/\sigma_n)$ of 0.82 will 
correspond to an inclination angle $\alpha$ = 76\deg, 60\deg and 36\deg\  
for $\Delta\theta$=20\deg, 45\deg and 90\deg\  respectively.
The dependence of the estimates for $\alpha$ on $\Delta\theta$ is less critical 
for $\Delta\theta \sim$ 90\deg, a case when turbluent motions dominate over 
collimated flows. 
Although MM1 in DR21(OH) does power high velocity CO outflows (Lai et al.\ 
2003), 
there is no evidence that the spectra of ions and neutrals observed 
at the rest velocity of the cores are contaminated by collimated   flows. 
Therefore, to derive the magnetic field inclination, we have 
ignored collimated flows and we assume the gas is fully 
turbulent ($\Delta\theta=90$\deg).
Figure 4 shows the inclination angle $\alpha$ ranges from 30\deg\ 
in the northern part of MM1 to 42\deg\ in the south, and the uncertainty in
$\alpha$ is $\sim$ 4--8\deg.  The variation in $\alpha$ is small (12\deg) and
is comparable to the variation of 
the overlaid $B_p$ vectors derived from the dust polarization
(black) and CO polarization (white).
Therefore, the magnetic field directions seem to be approximately
uniform across the MM1 region.  
The average line-of-sight magnetic field strength in
MM1 $\sim$ $-$0.36$\pm$0.10 mG as measured by 
CN Zeeman observations (Crutcher et al.\ 1999).
With the average $\alpha=36$\deg, 
the field strength in the plane of sky ($B_p$) is 0.25$\pm$0.07 mG, and
the full magnetic field strength is $ 0.44\pm$0.12 mG. The direction of the 
field is out of the plane of the sky towards the observer at an angle of 
36\deg to the line of sight, and its projection on the sky is at a 
position angle $-75$\deg. 

Lai et al.\ (2003) estimate $B_p\sim$0.9 mG for MM1 from the observed 
dispersion in the CO polarization position angles using
the Chandrasekhar-Fermi formula.
This value is a factor of 3 larger than what is derived from the cyclotron 
interaction data presented here. This discrepancy 
could be partly due to the low values for the dispersion caused by the 
smoothing of polarization by the finite beam size of the BIMA observations.
Also, the `calibration factor' of the Chandrasekhar-Fermi formula (0.5)
used by Lai et al.\ (2003) is an average value of clouds
in numerical simulations (Ostriker, Gammie, \& Stone 2001), 
which may not be a proper value DR21(OH).  
 
With the knowledge of the full magnetic field strength, we can evaluate 
the role that magnetic fields play for star formation in MM1.
First, we can calculate the mass-to-magnetic flux ratio
from $M/\Phi_B=1.0\times 10^{-20}N(H_2)/|B|$
in  units of the critical mass-to-flux ratio (Crutcher 1999),
where $N(H_2)$ is the column density in cm$^{-2}$
and $|B|$ is the field strength in $\mu$G.
With $N(H_2)=2\times 10^{23}$ cm$^{-2}$ 
and B=0.44 mG, $M/\Phi_B=$4.5.
Therefore, MM1 is clearly magnetically supercritical.
However, turbulence seems to provide dominant support.
The H$^{13}$CO$^+$ linewidth ($\sim$ 2.2\kms) is much
larger than its thermal linewidth for $T_k$=50 K (0.3\kms),
indicating strong turbulence in DR21(OH).
The relative importance of magnetic, thermal, and turbulent support
can be evaluated by comparing their energy to the gravitational
energy.
We find that in MM1 the magnetic-to-gravitational energy ratio
is 0.05, the thermal-to-gravitational energy ratio is 0.16,
and the turbulent-to-gravitational energy ratio is 0.66.   
Therefore, turbulence provides the dominant support in MM1.  
Although our conclusion on the status of MM1 is similar to
what has been suggested by Crutcher et al.\ (1999)
with statistical argument for the full magnetic field strength,
our results exclude the possiblity of the magnetic field dominance
which may be true only if the field orientation in 3-D is very close to 
the plane of the sky.

\acknowledgements
We thank the referee for helpful suggestions. This research was conducted at the 
Jet Propulsion Laboratory, California Institute of Technology, with support from 
the National Aeronautics and Space Administration,
while SPL held a National Research Council Resident Research Associateship
at JPL. The OVRO Millimeter  Array is supported by NSF grant AST-99-81546.

\begin{figure} 
\plotone{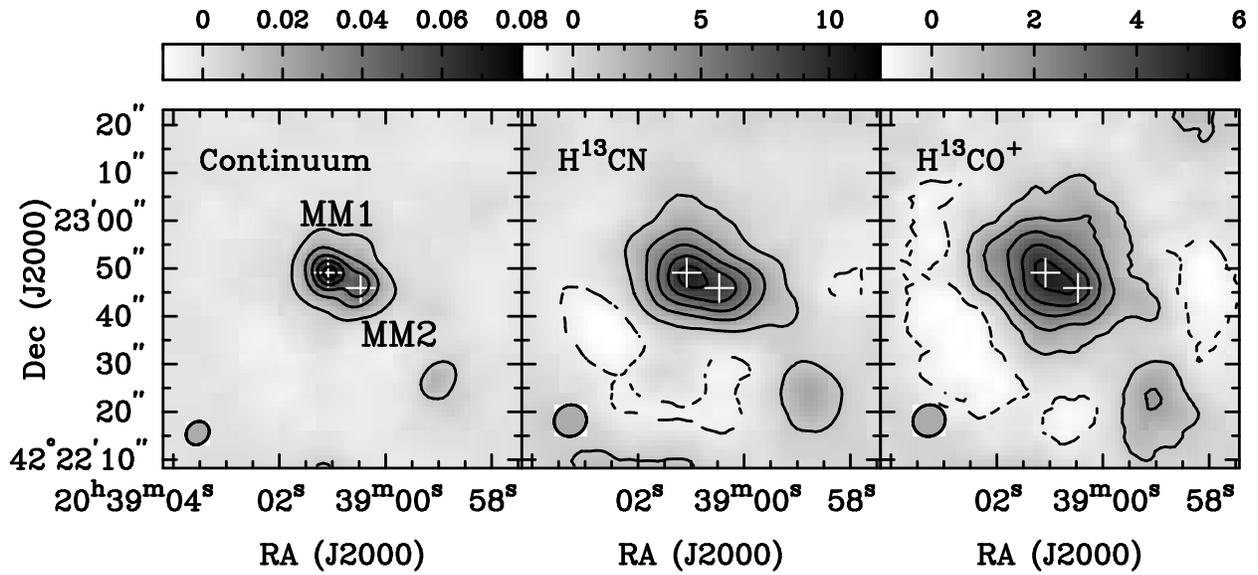}
\caption{The 3mm dust continuum and the velocity integrated intensity maps of
H$^{13}$CN and H$^{13}$CO$^+$. The lowest contour   and the interval are at   
10\% and 20\%, respectively, of the peak intensities.}
\end{figure}

\begin{figure} 
\plotone{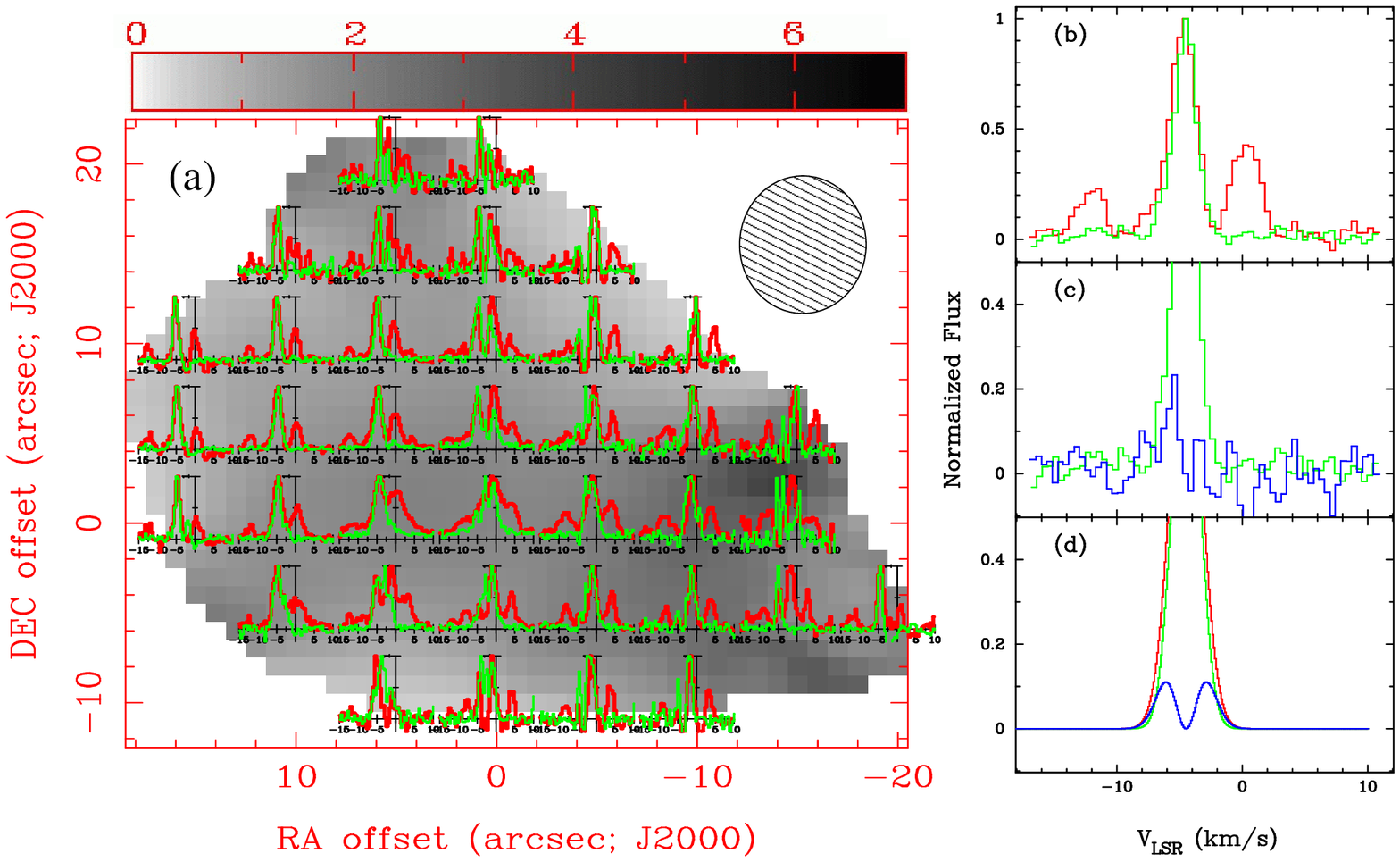}
\caption{Spectra of Ions and Neutrals. 
{\bf (a)} Normalized H$^{13}$CN  (red) and H$^{13}$CO$^+$ (green)
spectra  overlaid on H$^{13}$CN-to-H$^{13}$CO$^+$ flux ratio
(greyscale). The beam size is indicated. {\bf (b)}  
H$^{13}$CN (red) and H$^{13}$CO$^+$  
(green) spectra at offset position (+10$^{\prime\prime}$, +5$^{\prime\prime}$).
{\bf (c)} Difference
spectrum at this position between observed H$^{13}$CN and simulated H$^{13}$CN 
(using 
H$^{13}$CO$^+$ profile as a template - see  \S 2) is shown in blue.   
H$^{13}$CO$^+$ spectrum 
(green) is shown for comparison. {\bf (d)} Difference
spectrum  for model Gaussian spectra with  $(\sigma_i / \sigma_n)$ = 0.86,  
equivalent to that for the (+10$^{\prime\prime}$, +5$^{\prime\prime}$) 
position.}
\end{figure}

\begin{figure} 
\plotone{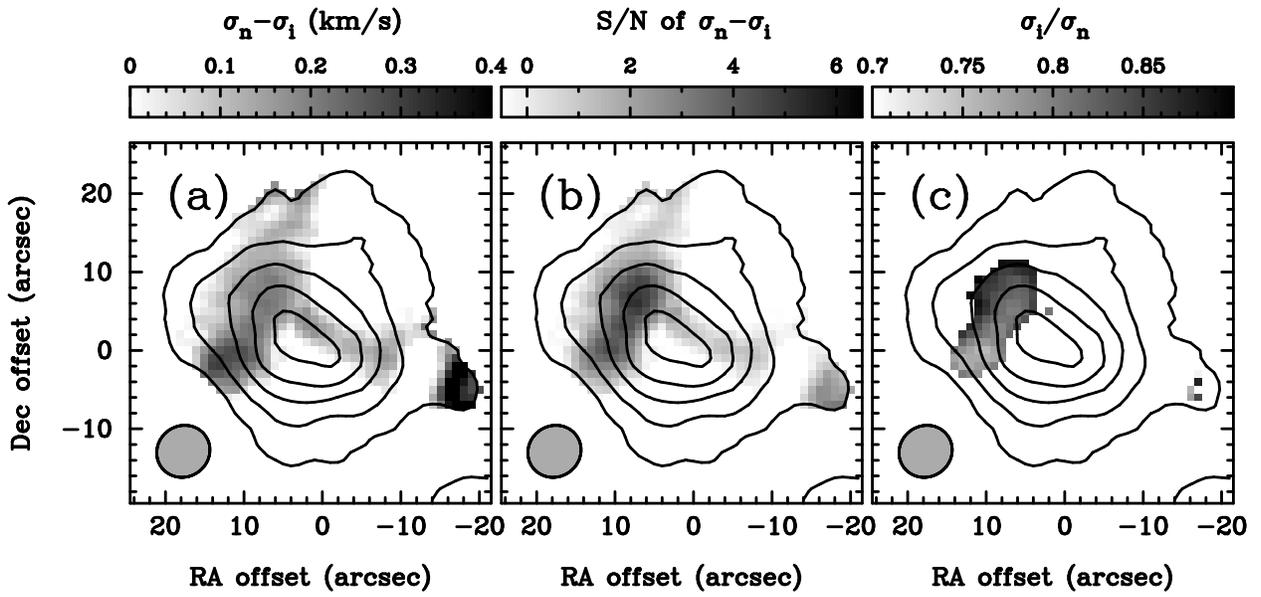}
\caption[]{Greyscale maps of ion/neutral linewidth differences overlaid on 
contours of  H$^{13}$CO$^+$ integrated flux at  10\%, 30\%, 50\%, 70\%, 
and 90\% levels. {\bf (a)} Difference of H$^{13}$CN and 
H$^{13}$CO$^+$
line dispersion $(\sigma_n - \sigma_i)$. {\bf (b)} Signal-to-noise
ratio of $(\sigma_n - \sigma_i)$. {\bf (c)}  Linewidth ratio $(\sigma_i / 
\sigma_n)$
for S/N of $(\sigma_n - \sigma_i) >$ 2.5. }
\end{figure}

\begin{figure} 
\epsscale{0.6}
\plotone{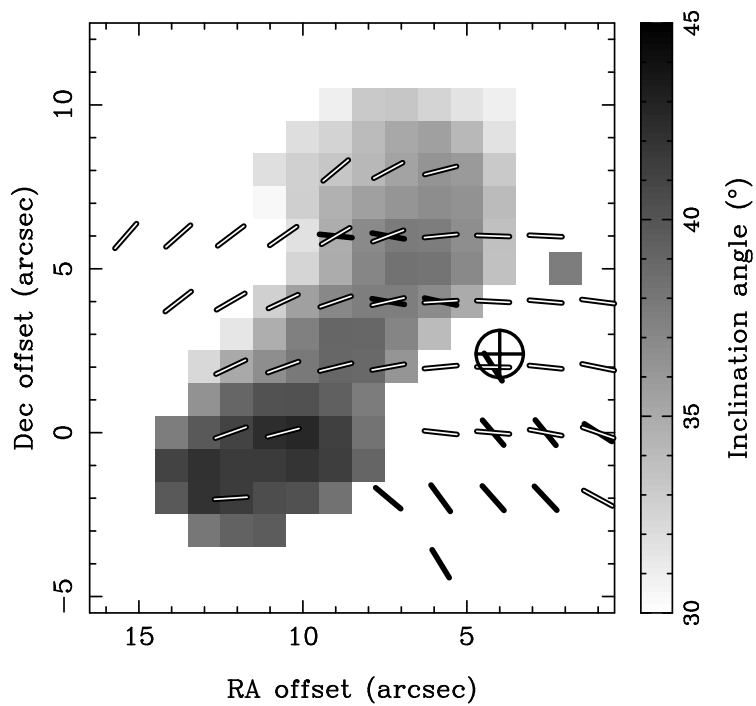}
\caption[]{Magnetic Field structure in DR21(OH) MM1 core. The greyscale 
shows the inclination angle of the magnetic field to 
the line of sight ($\alpha$)  varying from 30\deg\ to 
42\deg. The vectors represent
$B_p$ orientation inferred from dust (black) and CO(white)  polarization.
$\oplus$  represnts  the CN Zeeman measurement of line-of-sight  component, 
$\sim 0.36$mG towards observer.   }
\end{figure}
\end{document}